\documentclass[prl,twocolumn]{revtex4}
\usepackage{appendix}
\usepackage{amssymb,amsmath}
\usepackage{graphicx}
\usepackage{float}
\usepackage{placeins}

\DeclareMathOperator{\sech}{sech}

\begin{document}

%%%%%%%

\title{Solitary Matter Waves in Combined Linear and Nonlinear Potentials: 
Detection, Stability, and Dynamics}

\author{Scott Holmes}
\affiliation{School of Physics and Astronomy, University of Birmingham, Birmingham, UK}

\author{Mason A. Porter}
\affiliation{Oxford Centre for Industrial and Applied Mathematics, 
Mathematical Institute, University of Oxford, Oxford, UK}

\author{Peter Kr\"uger}
\affiliation{Midlands Ultracold Atom Research Centre, School of Physics \& Astronomy, The University of Nottingham, Nottingham, UK}

\author{Panayotis\ G.\ Kevrekidis}
\affiliation{Department of Mathematics and Statistics, University of Massachusetts,
Amherst, MA, USA}

%%%%%%%%%

%%%%%%%%%%

\begin{abstract}

We study statically homogeneous Bose-Einstein condensates with spatially inhomogeneous interactions and outline an experimental realization of compensating linear and nonlinear potentials that can yield constant-density solutions. We illustrate how the presence of a step in the nonlinearity coefficient can only be revealed dynamically and consider, in particular, how to reveal it by exploiting
the inhomogeneity of the sound speed with a defect-dragging experiment. We conduct computational experiments and observe the spontaneous emergence of 
dark solitary waves. We use effective-potential theory to perform a detailed analytical investigation of the existence and stability of solitary waves in this setting, and we corroborate these results computationally using a Bogoliubov-de Gennes linear stability analysis.  We find that dark solitary waves are unstable for all step widths, whereas bright solitary waves can become stable through a symmetry-breaking bifurcation as one varies the step width. Using phase-plane analysis, we illustrate the scenarios that permit this bifurcation and explore the dynamical
outcomes of the interaction between the solitary wave and the step.

\end{abstract}

%%%%%%%%%

%%%%%%%%%%%%%

\maketitle

%%%%%%%%%%%

%%%%%%%%%%%

\section{Introduction}

For more than two decades, Bose-Einstein condensates (BECs) have provided a fruitful experimental, computational, and theoretical testbed for investigating nonlinear phenomena.  In the mean-field limit, a BEC is governed by the Gross-Pitaevskii (GP) equation \cite{book1}, which is a nonlinear Schr\"odinger (NLS) equation with an external potential.  The NLS equation is important in many fields \cite{Sul}, and many ideas from disciplines such as nonlinear optics have proven important for investigations of BECs.  Moreover, the ability to control various parameters in the GP equation makes it possible to create a wide range of nonlinear excitations, and phenomena such as bright \cite{B1,B2}, dark \cite{D1,D2,D3}, and gap \cite{G} solitary waves (and their multi-component \cite{MC} and higher-dimensional \cite{Pots1,Pots2} generalizations) have been studied in great detail using a variety of external potentials \cite{Pots1,Pots2}.  

The GP equation's cubic nonlinearity arises from a BEC's interatomic interactions, which are characterized by the $s$-wave scattering length.  The sign and magnitude of such interactions can be controlled using Feshbach resonances \cite{Koehler,feshbachNa,ofr}, and this has led to a wealth of interesting theoretical and experimental scenarios \cite{theor,B1,B2,exp}.  In a recent example, Feshbach resonances were used to induce spatial inhomogeneities in the scattering length in Yb BECs \cite{Tak}. Such {\it collisional inhomogeneities}, which amount to placing the BEC in a nonlinear potential in addition to the usual linear potential, can lead to effects that are absent in spatially uniform condensates~\cite{NLPots,Chang,Summary}.  This includes adiabatic compression of matter waves \cite{our1}, enhancement of the 
transmission of matter waves through barriers \cite{our2}, dynamical trapping 
of solitary waves \cite{our2}, delocalization transitions of matter 
waves \cite{LocDeloc}, and many other phenomena.  Nonlinear potentials have also led to interesting insights in studies of photonic structures in optics \cite{kominis}.

In the present paper, we study the situation that arises when spatial inhomogeneities in nonlinear and linear potentials are tailored in such a way that they compensate each other to yield a constant-density solution of the GP equation. We demonstrate how to engineer this scenario in experiments and investigate it for a step-like configuration of the potentials. This situation is particularly interesting because the inhomogeneity 
is {\it not} mirrored in the BEC's density profile, which makes the
step indistinguishable from a homogeneous linear and nonlinear potential in {\em static} density measurements.  We show that the step is nevertheless revealed {\em dynamically} in an impurity-dragging
experiment~\cite{bpa_imp}, and we observe the emission of dark solitary waves when the dragging speed is above a critical velocity (which is different inside
and outside of the step).  This spontaneous emergence of solitary waves
motivates their study as a dynamical entity
in this setting. We use effective-potential theory to examine the existence and potential
dynamical robustness of dark and bright quasi-one-dimensional (quasi-1D) solitary waves for various step-potential parameters. We find that dark solitary waves are always dynamically unstable as stationary states inside of the step, although the type of their instability varies depending
on the step parameters. In contrast, bright solitary waves experience a symmetry-breaking bifurcation as the step width is increased, so we analyze their dynamics using a
phase-plane description of their motion through the step.
Our effective-potential picture enables not only the unveiling of interesting
bifurcation phenomena but also an understanding of the potential dynamical
outcomes of the interaction of solitary waves with such steps.

In this paper, we highlight the fundamental difference between linear and nonlinear potentials in the dynamics of a quantum degenerate one-dimensional Bose gas. In the static picture, one type of potential can be adjusted to completely compensate the other, so that there is no difference to the simple homogeneous potential landscape. However, the dynamical picture is different, as a flow of the Bose gas across inhomogeneities displays interesting dynamics.  In the present investigation, we use step potentials to illustrate this phenomenon.

The remainder of this paper is organized as follows.  We first present our model and its associated physical setup.  We then discuss a proposal for the
experimental implementation of the compensating linear and nonlinear potentials that we discussed above. We then discuss the problem of dragging a moving defect through the step and the ensuing spontaneous emergence of solitary waves.  We then examine the existence, stability, and dynamics of the solitary waves both theoretically and computationally. Finally, we summarize our findings and propose several directions for future study.

%%%%%%%%%%

\section{Model and Setup} 

\begin{figure}[ht]
\centering
\includegraphics[width=0.5\textwidth,clip,trim = 0 5 0 6]{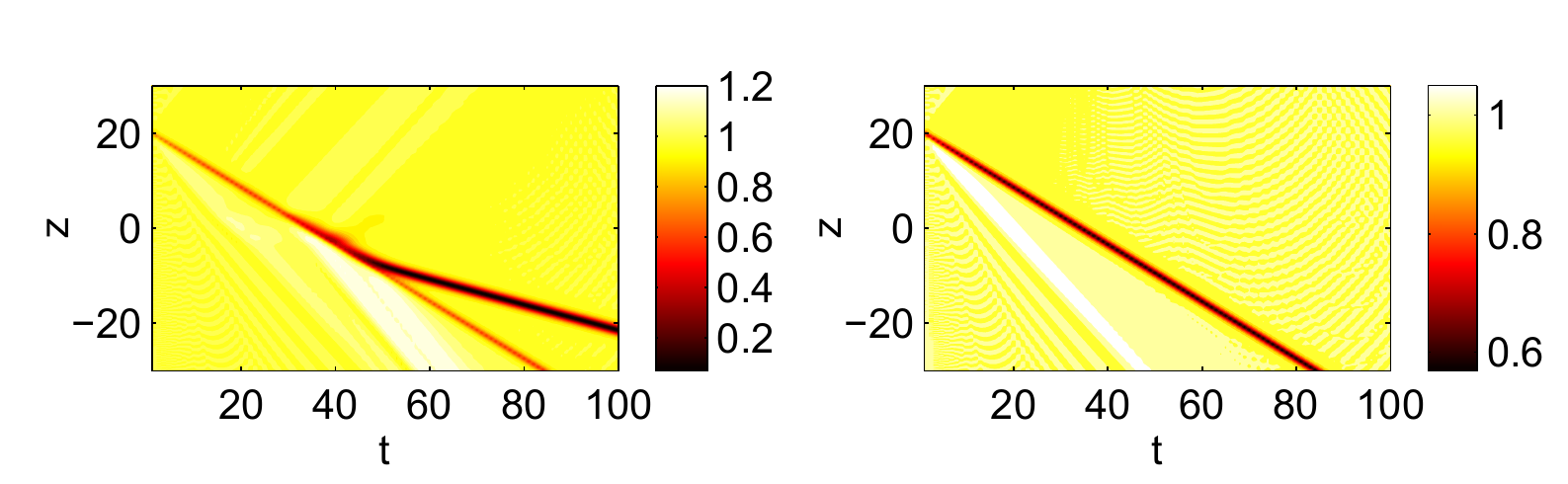}
\caption{[Color online] Numerical computations of defect dragging in the quasi-1D GP equation.  (Left) Emission of a dark solitary wave as a defect is dragged through the step.  (Right) The same computational experiment but without a step (so there is no solitary-wave emission).  The defect speed is \(v=0.6\), and the other parameter values are \(\gamma = -1\) and \(\Delta V = 0.5\).}
\label{fig-dd}
\end{figure}

We start with the three-dimensional (3D) time-dependent GP equation and consider a cigar-shaped condensate by averaging over the transverse directions to obtain a quasi-1D GP equation \cite{book1,Pots1,Pots2}.  In performing the averaging, we assume that the BEC is strongly confined in the two transverse directions with a trapping frequency of \(\omega_\perp\)~\footnote{Our considerations can be extended to more accurate quasi-1D models \cite{accurate}, but we employ the quasi-1D GP setting for simplicity.}. The solution of the quasi-1D GP equation is a time-dependent macroscopic wavefunction $\Psi(z,t)$. We use the standing-wave ansatz \(\Psi(z,t) = \phi(z)e^{-i\mu t}\) to obtain the time-independent GP equation
\begin{equation}\label{gpe}
	-\frac{1}{2}\phi_{zz} -\mu \phi + V_{\mathrm{ext}}(z)\phi + g(z)\vert\phi\vert^2\phi=0\,,
\end{equation}
where \(\phi\) is measured in units of \((2\vert a_0\vert)^{-1/2}\) and 
\(g(z)\) is a spatially varying nonlinearity associated with the 
(rescaled) scattering length \(a(z)\) via \(g(z) = a(z)/\vert a_0\vert\). We measure length in units of \(a_\perp \equiv \sqrt{\hbar/(m\omega_\perp)}\) and time in units of $\omega_\perp^{-1}$, where \(m\) is the mass of the atomic species forming the condensate. The constant \(a_0\) is the value of the scattering length in the collisionally homogeneous system.
 Equation (\ref{gpe}) has two conserved quantities: the number of atoms \(N = (a_\perp/[2\vert a_0\vert])\int^{+\infty}_{-\infty}\vert\Psi\vert^2 dz\)  and the Hamiltonian \cite{Pots2}. 

For a square-step linear potential, one can use the Thomas-Fermi 
approximation ($\phi_{zz} = 0$) for the ground state \cite{Pots2}. Equating the densities inside and outside of the step then gives the constraint
\begin{equation} \label{gamma}
	\gamma = \frac{\Delta V}{\Delta g}=\frac{V_0-\mu}{g_0}\,, 
\end{equation}
where \(V_{0}\) and \(g_{0}\) are the constant background linear and nonlinear potentials, and \(\Delta V\) and \(\Delta g\) are the differences between the step and background values of \(V(z)\) and \(g(z)\).  The parameter \(\gamma\) thus measures (and balances) the relative strengths of the steps in the linear and nonlinear potentials.  To preserve smoothness, we implement the steps using hyperbolic tangent functions:
\begin{align}
	\!\!\!V(z) \!&=\! V_0 +\Delta V(z) = V_0 \!+\! \frac{\Delta V}{2}\left[\tanh(z_+)-\tanh(z_-)\right], \notag \\
	\!\!\!g(z) \!&=\! g_0 + \Delta g(z) = g_0 \!+\! \frac{\Delta g}{2}\left[\tanh(z_+)-\tanh(z_-)\right],
\end{align} 
where \(z_{\pm} = (z\pm z_0)/s\), the step width is $2z_0$, and \(s\) controls the sharpness of the step edges.  From equation (\ref{gamma}), it follows that \(\Delta V = \gamma \Delta g\).  For the remainder of this article, we take \(V_0=0\) and \(\vert g_1\vert=\vert\mu\vert=1\).  This yields \(\gamma = -1\) and corresponds to nonlinear and linear steps of equal and opposite depths/heights.

%%%%%%%%

\section{Proposal for Experimental Implementation}

Techniques for manipulating cold quantum gases have become both advanced and accurate, and they allow experimentalists to form a variety of potentials with optical and/or magnetic fields, especially near microstuctured atom chips \cite{atomchiprev, chipNJP}. It was shown recently that spatially varying nonlinear potentials, which have been of theoretical interest for several years \cite{NLPots,Chang,Summary}, can be used address a novel scenario that can also be implemented experimentally \cite{Tak}. Straightforward implications of a spatial inhomogeneity of the coupling coefficient $g$ include static density variations as a result of the inhomogeneous mean field. To distinguish this type of effect from more subtle dynamical and beyond-mean-field phenomena, it is desirable to compensate linear and nonlinear contributions of the potential in such a way that the static density profile remains homogeneous (as would be the case if all potentials were homogeneous). In this section, we discuss how such a situation can be achieved experimentally. (In the next section, we will give an example of a purely dynamical phenomenon that arises from it.)
 
A spatially varying magnetic field $B(z)$ results in a proportionally varying linear potential $V(z)=m_F g_F \mu_B B(z)$ for magnetic spin states (where the magnetic quantum number is $m_F$, the Land\'{e} factor is $g_F$, and the Bohr magneton is $\mu_B$) at sufficiently low magnetic fields within the regime of validity of the linear Zeeman effect. For specific atomic species and spin states, there is an additional resonant dependence (a Feshbach resonance \cite{FeshRev}) of $g$ on the magnetic field: 
\begin{equation}
	g(B)=g_\mathrm{bg}\left(1-\frac{\Delta}{B-B_0}\right)\,, 
\end{equation}	
where $g_\mathrm{bg}$ is the background coupling constant, $B_0$ is the resonance field, and $\Delta$ is the resonance width. The condition of 
compensating linear and nonlinear potentials is fulfilled within the Thomas-Fermi approximation when 
\begin{equation}\label{req}
	n \frac{\partial g}{\partial B}=-\frac{\partial V}{\partial B}\,. 
\end{equation}	
In theory, this implies for any given density $n$ that there is a field $B_c$ near the resonance $B_0$ that satisfies equation (\ref{req}). Consequently, the density must remain constant for any static profile $B(z)$ as long as $B(z) - B_c$ is sufficiently small (so that $g(B)$ is an approximately linear function of $B$). 

In practice, however, large nonlinearities lead to fast three-body recombination losses from traps and hence have to be avoided \cite{FeshRev}. An atomic species with appropriate properties is cesium, for which the above conditions are fulfilled at typical densities of $10^{13}-10^{14}$ cm$^{-3}$ for fields near the narrow Feshbach resonances at 19.8 G and 53.5 G \cite{ChinCs}. 

Optical dipole traps near the surface of atom chips \cite{chipdipole} provide an environment in which magnetic fields can be accurately tuned to and varied about the critical magnetic fields $B_c$ at the above parameter values. One can bring the trap close to independent microstructures on the surface of the chip by coating the surface with a highly reflective layer so that a standing light wave forms a 1D optical lattice whose near-surface wells can be loaded with the atomic sample.  Alternatively, one can focus a single laser beam to a position near the surface at a frequency that is slightly below that of the main atomic transition (i.e., one can red-detune it). In this case, integrated optics and microlenses might help to reduce the atom-surface distance $d_\mathrm{surf}$ to the single-micron regime. Once the trap is placed and populated with an atomic sample, currents that pass through appropriately shaped surface-mounted conductor patterns produce the necessary magnetic field profiles that we described above. The field-tailoring resolution and hence the width of a possible step is limited by $d_\mathrm{surf}$. It is feasible to reduce this length to roughly $1\mu$m in current experiments.  In particular, one can exploit the lattice approach \cite{chipdipole}, in which the closest wells form at $d_\mathrm{surf}\approx \lambda$, where the wavelength $\lambda$ is in the optical range (i.e., $\lambda\lesssim 1\mu$m).

%%%%%%%%%%%

\section{Dragging a Defect Through the Step} 

Using the above techniques, the effect of a step on the static denisty profile can be removed by construction. In this case, it is interesting to investigate if and how the density profile is modified when a step is moving relative to the gas. We show by performing computational experiments that the presence of steps in the linear and nonlinear potentials can be revealed by dragging a defect through the BEC \cite{bpa_imp,hakim}. For the linear and nonlinear steps that we described above, the condensate density is constant within and outside of the step. However, the speed of sound $c$ is different in the two regions: 
\begin{equation}
	c = \sqrt{g(z)n(z)}\,, 
\end{equation}	
where \(n(z) = \vert\phi(z)\vert^2\) is the BEC density \cite{sound}.  To perform computations that parallel viable experiments, we simulate a moving defect using a potential of the form 
\begin{equation}
	V(z,t) = Ae^{-[z-r(t)]^2/w^2}\,, 
\end{equation}	
where \(r(t) = r(0) + vt\) represents the center of a defect that moves with speed \(v\) and $A$ and $w$ are amplitude- and width-related constants.  The dynamics of defects moving in a BEC are sensitive to the speed of the defect relative to the speed of sound: speeds in excess of the speed of sound (i.e., supercritical defects) lead to the formation of dark solitary waves travelling behind the defect, whereas speeds below the speed of sound (i.e., subcritical defects) do not~\cite{hakim}.  

There are three possible scenarios. First, when the speed is subcritical, there is a density depression with essentially the same functional form as the linear potential. This changes shape slightly in the presence of the step; it deepens and widens for a step with \(\Delta g < 0\), and it becomes shallower and narrower when \(\Delta g > 0\)~\footnote{Additionally, initialization of the moving step or an impact on the step produces small-amplitude, oscillatory Hamiltonian shock waves \cite{hoefer}.}. When the speed is larger but still subcritical, the situation is similar---except that the depression distorts slightly, giving rise to a density hump in front of the defect. Second, when the defect speed is supercritical within the step region but subcritical outside of it, we expect the nucleation of dark solitary waves in the step region.  Because the defect's speed is smaller than the background sound speed, the emission of solitary waves downstream of the defect becomes a clear indication of the presence of a step.  We demonstrate this scenario in Fig.~\ref{fig-dd}.  The third possible scenario involves a defect that is supercritical in both regions.

%%%%%%%%%%

\section{Existence, Stability, and Dynamics of Solitary Waves. Part I: 
Theoretical Analysis} 

Our scheme for applying compensating steps to the linear and nonlinear potentials and our ensuing observation that solitary waves emerge from moving steps warrant a detailed investigation of 
the dynamics in this scenario. In particular, we examine the existence and stability of solitary-wave solutions as a function of step parameters (especially step width).

%%%%%%

\subsection{Bogoliubov-de Gennes Analysis} 

We apply the Bogoliubov-de Gennes (BdG) ansatz
\begin{align}\label{linstab}
	\!\!\!\Psi(z,t) = e^{-i\mu t}\left[\phi_0(z) \!+\! \sum_j(u_j(z)e^{-i\omega_jt} \!+\! v^*_j(z)e^{i\omega_jt})\right]
\end{align} 
to the time-dependent quasi-1D GP equation.  Equation (\ref{linstab}) defines the linear eigenfrequencies \(\omega_j\) for small 
perturbations that are characterized by eigenvectors \(u_j(z)\) and \(v_j(z)\).  
Linearizing the time-dependent GP equation about the 
reference state \(\phi_0(z)\) using equation (\ref{linstab}) yields the BdG eigenvalue problem. 
The eigenfrequencies \(\omega_j\) come in real (marginally
stable) or imaginary (exponentially unstable) pairs or as
complex (oscillatorally unstable) quartets.

In our analytical approach, we examine perturbations of the 
time-independent GP equation (\ref{gpe}) with constant potentials $V(z) \equiv V_0 = 0$ and $g(z) \equiv g_0 =\pm1$.  The perturbations in the linear and nonlinear steps are thus \(\Delta g(z)\) and \(\Delta V(z) = \gamma\Delta g(z)\).  We introduce \(\epsilon \equiv \vert\Delta g\vert\) as a small parameter and (to facilitate presentation) use the term ``negative width" to describe a step with $\Delta g < 0$.  When $g_0=\pm 1$, equation (\ref{gpe}) has two families of (stationary) 
soliton solutions, which are characterized by center position $\xi$ and chemical potential $\mu$.  The case $g_0=-1$ yields bright solitons:
\begin{equation}
	\phi_{\mathrm{bs}}(z-\xi) = \eta_{\mathrm{bs}} \sech\left(\eta_{\mathrm{bs}}(z-\xi)\right)\,, 
\end{equation}	
where \(\eta_{\mathrm{bs}}=\sqrt{-2\mu}\) and \(\mu<0\).  The case $g_0=1$ yields dark solitons: 
\begin{equation}
	\phi_{\mathrm{ds}}(z-\xi) = \eta_{\mathrm{ds}} \tanh\left(\eta_{\mathrm{ds}}(z-\xi)\right)\,, 
\end{equation}
where \(\eta_{\mathrm{ds}}=\sqrt{\mu}\) and \(\mu>0\).

%%%%%%%%%%%%%

\subsection{Effective-Potential Theory} 

We use a Melnikov analysis to determine the persistence of bright~\cite{Sand} 
and dark solitary waves \cite{Pel}.  Stable (respectively, unstable) solitary waves exist at minima (respectively, maxima) of an effective potential $M_{\mathrm{bs}}$. We find that 
bright solitary waves can, in principle, be stable within the step in the potentials. However, in contrast to the bright solitary waves, stationary dark solitary waves are generically unstable within the step.

To determine the persistence of a bright solitary wave, we calculate when its center position induces its associated Melnikov function (i.e., perturbed energy gradient)~\cite{Sand} to vanish.  This yields the equation
\begin{align}\label{melbright}
	M_{\mathrm{bs}}'(\xi_0) &= \int_{-\infty}^\infty \biggl[\frac{d[\Delta V(z)]}{dz}\phi_{\mathrm{bs}}^2(z-\xi_0) \notag \\
	&+ \frac{1}{2}\frac{d[\Delta g(z)]}{dz}\phi_{\mathrm{bs}}^4(z-\xi_0)\biggr] dz = 0
\end{align}
for the first derivative of the potential at the solitary-wave center $\xi = \xi_0$.

The GP equation without a potential is spatially homogeneous, and it possesses translational and $U(1)$-gauge symmetries. These symmetries are associated with a quartet of eigenfrequencies at the origin.  When the translational symmetry is broken (e.g., by the steps in \(V(z)\) and \(g(z)\)), a pair of eigenfrequencies leaves the origin. Tracking their evolution makes it possible to examine the stability of solitary waves of the perturbed system.  We follow these eigenfrequencies by computing the function  
\begin{align}
	M_{\mathrm{bs}}''(\xi_0) &= \int_{-\infty}^\infty \biggl[\frac{d^2[\Delta V(z)]}{dz^2}\phi_{\mathrm{bs}}^2(z-\xi_0) \notag \\
 &+ \frac{1}{2}\frac{d^2[\Delta g(z)]}{dz^2}\phi_{\mathrm{bs}}^4(z-\xi_0)\biggr] dz\,,
\end{align}
which determines the concavity of the perturbed energy landscape and is directly associated to the eigenfrequencies of the linearization through~\cite{Sand}
\begin{equation}
	\omega^2 = \frac{1}{2\sqrt{-2\mu}}M_{\mathrm{bs}}''(\xi_0) + O(\epsilon^2)\,,
\end{equation} 
where we note that $M_{\mathrm{bs}}''(\xi_0) = O(\epsilon)$. Stable (respectively, unstable) solitary waves exist at minima (respectively, maxima) of the effective potential $M_{\mathrm{bs}}$. Hence, bright solitary waves can, in principle, be stable within the step. 

We compute analogous expressions for dark solitary waves, but the Melnikov function now needs to be renormalized due to the presence of a nonzero background density~\cite{Pel}.  The first and second derivatives of the effective potential $M_{\mathrm{ds}}$ evaluated at the solitary-wave center $\xi = \xi_0$ are
\begin{align}
	M_{\mathrm{ds}}'(\xi_0) &= \int_{-\infty}^\infty \biggl[\frac{d[\Delta V(z)]}{dz}\left[\eta_{\mathrm{ds}}^2-\phi_{\mathrm{ds}}^2(z-\xi_0)\right]\nonumber \\
	&+ \frac{1}{2}\frac{d[\Delta g(z)]}{dz}\left[\eta_{ds}^4- \phi_{\mathrm{ds}}^4(z-\xi_0)\right]\biggr] dz =0
\end{align}
and 
\begin{align}
	M_{\mathrm{ds}}''(\xi_0) &= \int_{-\infty}^\infty \biggl[\frac{d^2[\Delta V(z)]}{dz^2}\left[\eta_{\mathrm{ds}}^2-\phi_{\mathrm{ds}}^2(z-\xi_0)\right] \nonumber \\
	&+ \frac{1}{2}\frac{d^2[\Delta g(z)]}{dz^2}\left[\eta_{\mathrm{ds}}^4- \phi_{\mathrm{ds}}^4(z-\xi_0)\right]\biggr] dz \neq 0\,.
\end{align}
The expression for the associated eigenfrequencies in this case is~\cite{Pel}
\begin{equation}
	\omega^2 = \frac{1}{4}M_{\mathrm{ds}}''(\xi_0)\left(1-\frac{i\omega}{2}\right) + O(\epsilon^2)\,,
\end{equation} 
where we choose the root that satisfies \(\mathrm{Re}(i\omega)>0\) and we note that $M_{\mathrm{ds}}''(\xi_0) = O(\epsilon)$.  

The main difference between the spectra for dark and bright solitary waves is that the continuous spectrum associated with the former (due to the background state) lacks a gap about the origin.  Consequently, exiting along the imaginary axis is not the only way for eigenfrequencies to become unstable. Even when eigenfrequencies exit toward the real axis, they immediately 
leave it as a result of their collision with the continuous spectrum; this leads to an eigenfrequency quartet. Thus, stationary  dark solitary waves are generically unstable within the step.

%%%%%%

\section{Existence, Stability, and Dynamics of Solitary Waves. Part II: 
Computational Results}

\begin{figure}[h!t]
\centering
\includegraphics[width=0.5\textwidth,clip,trim = 0 6 0 5]{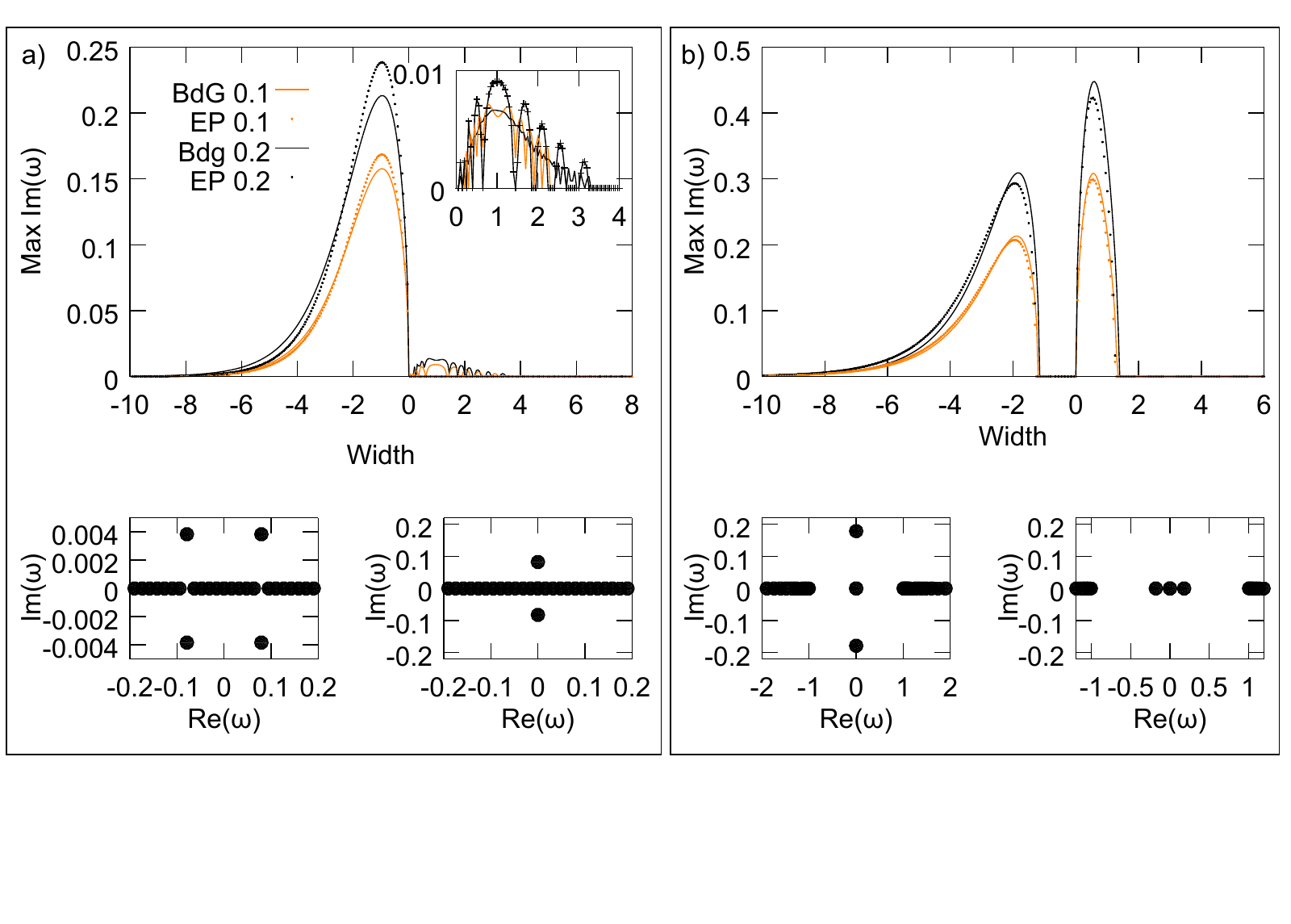}
\vspace{-1.5cm}
\caption{[Color online] (Top) Maximum imaginary eigenfrequencies versus step width 
(where a negative step width means that \(\Delta g < 0\)) 
for (left) dark solitary waves and (right) bright solitary waves.  
We show results for the perturbation strengths \(\epsilon = 0.1\) and \(\epsilon = 0.2\).  Dashed curves give results for analytical calculations from effective-potential theory, and solid curves give numerical calculations using the BdG equations. The inset in the left panel shows 
finite-size effects (see the main text). (Bottom) Examples of the corresponding eigenfrequency spectra for $\epsilon = 0.1$.  For both bright and dark solitary waves, we show the spectrum for a step width of $2z_0 = 0.25$ on the left and a step width of $2z_0 = -0.25$ on the right.
%{\bf map: pink vs orange in inset: can this be distinguished in grayscale?}
%{\bf also: please send me a version of this with \emph{axes} (I finally figured out what the journal had in mind!  none of the figures here have the axes indicated, so let's put those in and see what they look like}
}
\label{fig-eigs}
\end{figure}

We use a fixed-point iteration scheme to identify stationary solitary-wave solutions, solve 
the BdG equations numerically to determine their corresponding eigenfrequencies, and employ parameter continuation to follow the solution branches as we vary the step width.  

We start with the $\xi_0 = 0$ branch, which exists for all step widths.
In Fig.~\ref{fig-eigs}, we show the development of the eigenfrequencies of 
this branch of solutions as a function of step width for both dark 
(left) and bright (right) solitary waves.  We obtain 
good \emph{quantitative} agreement between our results from effective-potential theory and those from BdG computations for the nonzero eigenfrequency associated with the intrinsic (translational)
dynamics of the solitary wave.

For the case of repulsive BECs ($g > 0$), the branch of solutions at $\xi = 0$ has a real instability for $\Delta g <0$ (i.e., $\Delta V > 0$) and an oscillatory instability for $\Delta g > 0$. We capture both types of instability accurately using effective-potential theory. An interesting but unphysical feature 
of the dark solitary waves is the presence of small ``jumps'' in the eigenfrequencies. 
These jumps are finite-size effects that arise from the discrete numerical approximation to the model's continuous spectrum \cite{Joh}. 

The case of attractive BECs ($g < 0$) is especially interesting. A pitchfork 
(symmetry-breaking) bifurcation occurs as the step widens; it is supercritical for \(\Delta g < 0\) and subcritical for \(\Delta g > 0\).  In this case, oscillatory instabilities are not possible when translational invariance is broken~\cite{Sand}. A direct and experimentally observable consequence of our analysis is that (for $\Delta g>0$) bright solitary waves remain stable for 
sufficiently large step width, whereas narrowing the step should eventually 
lead to unstable dynamics.  For dark solitary waves, by constrast, we expect the dynamics to be unstable in experiments for all step widths.

\begin{figure*}[h!t]
\centering
\includegraphics[width=1\textwidth,clip,trim = 0 5 0 5]{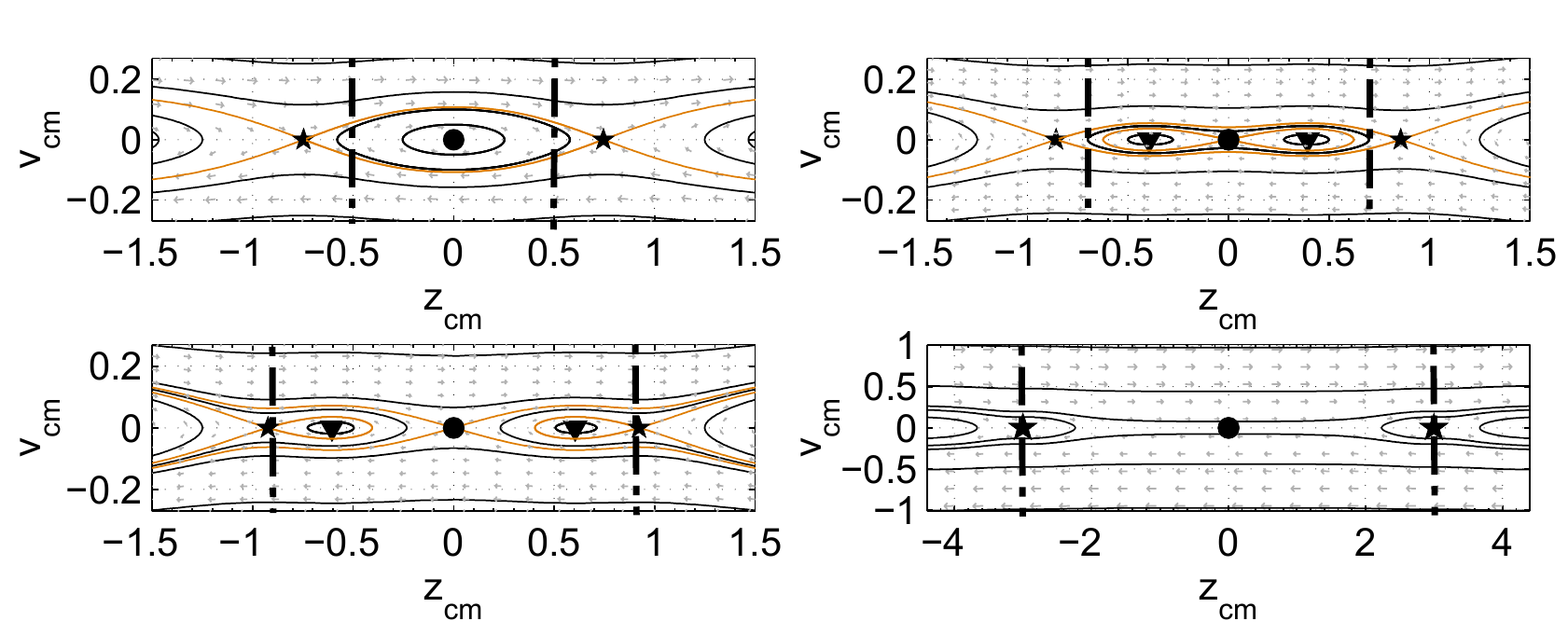}
\caption{[Color online] Phase planes for Newtonian dynamics that describe 
bright solitary waves in an attractive BEC for four different step widths. The thick dash-dotted lines represent the edges of the step. We highlight the equilibria with dots, triangles, and stars.  The light (orange) curves correspond to trajectories that originate at equilibria, and we show other example trajectories as dark (black) curves.  The step widths are (upper left) $2z_0 = -1$, (upper right) $2z_0 = -1.4$, (lower left) $2z_0 = -1.8$, and (lower right) $2z_0 = -6$.
}
\label{fig-pplane}
\end{figure*}

To further probe the bifurcation, we study the Newtonian dynamics~\cite{dyn}
of the bright solitary wave:  
\begin{equation}\label{newt}
	m_{\mathrm{eff}}\frac{d^2\xi}{dt^2}=-\nabla U(\xi) = 2M'_{\mathrm{bs}}(\xi)/N\,, 
\end{equation}	
where the effective mass is $m_{\mathrm{eff}} = 1/2$. We examine phase portraits of equation (\ref{newt}) by plotting the center-of-mass position $z_{\mathrm{cm}} \approx \xi$ versus the center-of-mass velocity $v_{\mathrm{cm}} \approx \frac{d\xi}{dt}$.  As we illustrate in Fig.~\ref{fig-pplane}, this is convenient for examining changes in dynamics as we alter the step width.  For narrow steps (e.g., a width of $2z_0 = -1$), there is a center at $z_{\mathrm{cm}}=0$ that straddles two saddle points (stars) just outside of the step (whose edges we indicate using dash-dotted lines). When $\Delta g < 0$ (i.e., $\Delta V >0$), a supercritical pitchfork bifurcation occurs at $2z_0 \approx -1.2$, as the center at the origin 
transitions to a pair of centers separated by a 
saddle at the origin (see the top right panel).  As the step widens further (bottom left panel), the heteroclinic orbit that previously enclosed the central three equilibria is
no longer present, and 
the centers are now surrounded by homoclinic orbits that emanate from the outer saddle points. Eventually, each outer saddle and its associated center 
annihilate one another (bottom right panel).  When $\Delta g > 0$, the types of equilibria are interchanged (saddles become centers and vice versa). The main difference that occurs in this case is that solitary waves can no longer be reflected by the step; they are all transmitted.  As one increases the magnitude of the step width from $0$, there is a saddle flanked by two centers.  At the bifurcation point, the central saddle splits into two saddles with a center between them.

%%%%%%

\begin{figure}[ht]
\centering
\includegraphics[width=0.5\textwidth,clip,trim = 0 4 0 8]{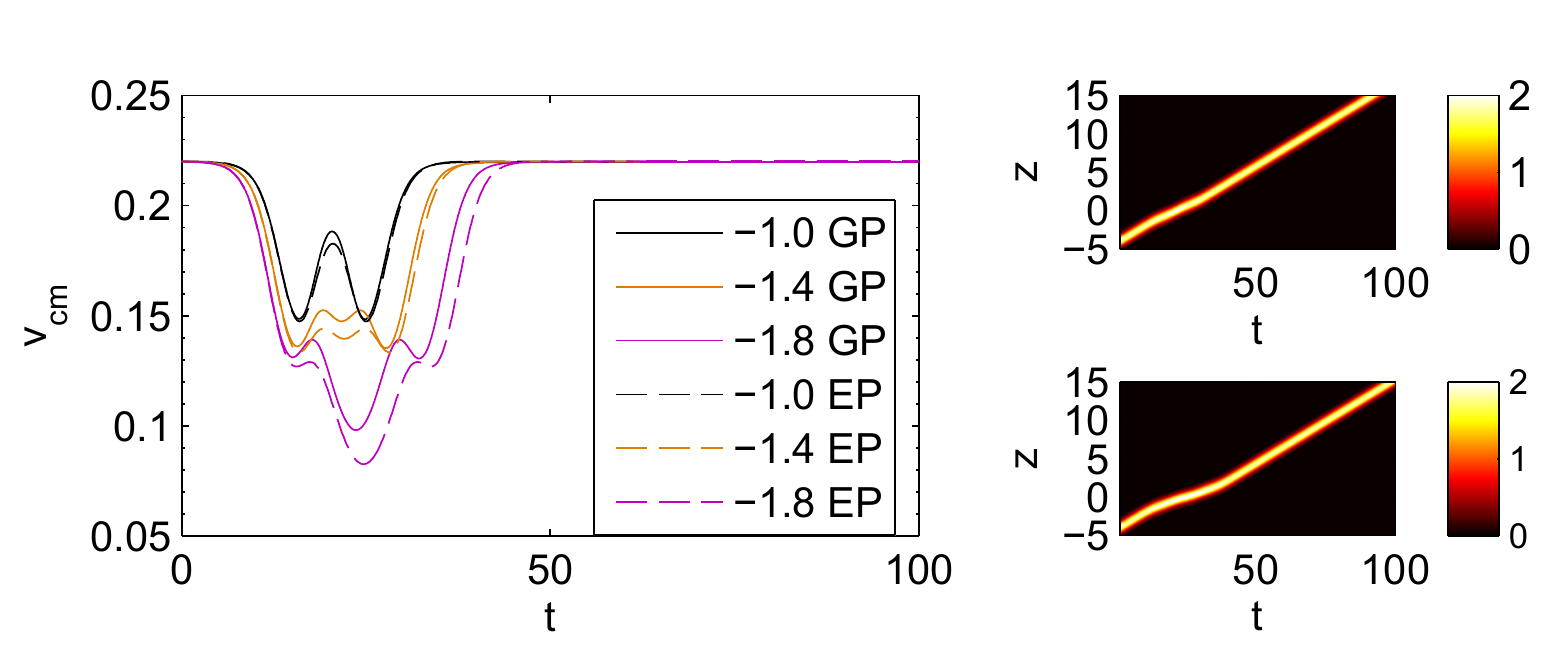}
\caption{[Color online] (Left) Effect of the step on the movement of a bright solitary wave for three different step widths for the GP equation (solid curves) and for numerical solutions of the Newtonian dynamics of the effective-potential (EP) equations (dashed curves). (Right) Contour plots of \(\vert\psi(z,t)\vert^2\) obtained by solving the GP equation numerically for step widths of (top) $-1$ and (bottom) $-1.8$.
%{\bf map: for Scott: can the orange vs pink be distinguished in grayscale?}
}
\label{fig-dyn}
\end{figure}

The changes to the possible trajectories in phase space suggest a viable way to investigate the bifurcation experimentally (and hence to distinguish between
narrow and wide steps).  The presence of a step alters the path of a moving solitary wave, as is particularly evident by examining the wave speed.  As we illustrate in Fig.~\ref{fig-dyn}, the solitary-wave dynamics depends on the number and type of phase-plane equilibria (and hence on the step width). The main panel shows how one can use variations in $v_{\mathrm{cm}}$ of a transmitted bright solitary wave to identify which equilibria are present. The center-of-mass motion of the solitary wave is a particularly useful quantity, as it is directly accessible to experimental measurement through time-resolved detection of spatial density profiles. The techniques outlined above for shaping the nonlinear potential---i.e., engineering the spatial profile $g(x)$ while automatically compensating the linear potential $V(x)$---gives a straightforward method to adjust the step width in the laboratory. 

We examine trajectories starting from the same initial conditions, $(z_{\mathrm{cm}}(0), v_{\mathrm{cm}}(0)) = (4,-0.22)$, for step widths of $-1,-1.4$, and $-1.8$.  The simplest trajectory occurs for the narrowest width ($2z_0 = -1$): as the solitary wave traverses the step, its speed first drops before rising again in the center of the step and then dropping again as it leaves the step (due to its encounter with the two saddles and the center in the phase plane; see Fig.~\ref{fig-pplane}).  For wider steps, the dynamics illustrate the effects of the bifurcation: instead of a single peak in the speed, there are now two peaks separated by a well. As the step widens further, the two peaks move outward and follow the centers to the edge of the step. The maximum and minimum in each pair move closer together in both $v_{\mathrm{cm}}$ and $t$ as one approaches the edge of the step. The solitary wave can either be transmitted (as illustrated in Fig.~\ref{fig-dyn}) or reflected by the step.

%%%%%%%%%%%%

\section{Conclusions} 

We introduced an experimentally realizable
setup to study statically homogeneous BECs in mutually compensating inhomogeneous linear and nonlinear potentials. We showed that---in contrast to the straightforward static scenario---a flowing gas will encounter sound-speed differences, which can induce interesting dynamics such as solitary-wave formation and motion. As a simple demonstration, we examined a step defect, whose width affects the system's dynamics. We conducted a thorough examination of solitary-wave stability and dynamics in this collisionally inhomogeneous setting.  We also showed how balancing linear and nonlinear potentials that yield constant-density solutions in the static case can be achieved experimentally.

We found that effective-potential theory gives a good \emph{quantitative} description of the existence and eigenfrequencies of both bright and dark solitary waves, and we used it to quantitatively track the evolution of the translational eigenfrequencies as a function of the step width.  We identified a symmetry-breaking bifurcation in the case of attractive BECs and illustrated how the presence of the bifurcation is revealed by the motion of solitary waves through the step region. We also found that stationary dark solitary waves are generically unstable through either exponential or oscillatory instabilities.

The system that we have studied provides a promising setup for future investigations, as it allows the experimentally realizable possibility of solitary-wave control via accurate, independent tailoring of linear and nonlinear potentials. It would be interesting to explore the phase-coherence properties of a collisionally inhomogeneous 1D quasicondensate, for which phase correlations (at $0$ temperature) decay algebraically with an interaction-dependent exponent \cite{1dpowerlaw}. Quasicondensates have comparitively small density fluctuations \cite{davis}.  In contrast to the scenario on which we have focused in the present paper, even a static quasicondensate gas would reveal a step in the nonlinearity in an interference experiment \cite{Kru2010} when the density profile is homogeneous. The study of such quasicondensates and of the phase fluctuations in them is a topic of considerable current interest~\cite{davis}, and it is desirable to enhance understanding of the properties of solitary waves in such systems.

%%%%%%%%%%

\section*{Acknowledgements} 

PGK acknowledges support from the US National Science Foundation (DMS-0806762),
the Alexander von Humboldt Foundation, and the Binational Science 
Foundation (grant 2010239). PK thanks the EPSRC and the EU for support.  We also thank an anonymous referee for helpful comments.

%%%%%%%%%%%%

%\newpage

\end{document}